\documentclass[11pt]{article} 
\usepackage{amsmath,amstext,amsbsy,amssymb}
\usepackage{bm}
\newcommand{\Tr}{{\rm Tr} }

\newcommand{\be}{\begin{equation}}
\newcommand{\ee}{\end{equation}}
\newcommand{\w}{\wedge}

\newcommand{\ssR}{{\scriptscriptstyle{R}}}

\newcommand{\ssS}{{\scriptscriptstyle{S}}}

\newcommand{\ssT}{{\scriptscriptstyle{T}}}

\newcommand{\ssW}{{\scriptscriptstyle{\rm W}}}

\newcommand{\ssM}{{\scriptscriptstyle{M}}}
\newcommand{\ssN}{{\scriptscriptstyle{N}}}

\long\def\symbolfootnote[#1]#2{\begingroup%
\def\thefootnote{\fnsymbol{footnote}}\footnote[#1]{#2}\endgroup}

\begin{document}


\begin{center}

{\Large \bf   
Topological Concepts for the Weyl Hamiltonians with the Berry Gauge Field 
}

\vspace{2cm}

\ Mahmut Elbistan

\vspace{5mm}

{\em {\it Physics Engineering Department, Faculty of Science and
Letters, Istanbul Technical University,\\
TR-34469, Maslak--Istanbul, Turkey}}\footnote{{\it E-mail address:} elbistan@itu.edu.tr }

\end{center}

\vspace{3cm}

The winding numbers for the even $d+1$ spacetime dimensional Weyl Hamiltonians are calculated in terms of the related Green's functions. It is shown that these winding numbers result in the divergence of the Dirac monopole fields, hence they are equal to the unit topological charge. It is demonstrated that the winding numbers are also equal to the Chern numbers which are expressed as the integral of the Berry field strength. Explicit calculations are presented for the $3+1$ and $5+1$ dimensional cases. Relevance of these topological invariants for the physical systems like the semiclassical chiral kinetic theory are discussed.

\vspace{11cm}

\pagebreak

\section{Introduction}
 
Physical systems may have nontrivial topological and geometrical properties. Besides their interesting mathematical structures, these properties yield important physical consequences. The simplest example is the magnetic monopole in the $3$ dimensional Euclidean space $\bm{R}^3$ where the existence of the monopole makes the space homotopic to the 2-sphere $S^2$. The gauge transformations on $S^2$ are classified by the fundamental group (or the first homotopy group) and this group is found to be isomorphic to the set of integers, $\Pi_1(S^1)\simeq \bm{Z}$. This integer is known as the winding number as it counts the number of times the gauge transformations wind around the gauge group $U(1)$. The winding number is also expressed in terms of the topological first Chern number. Another example that has numerous applications in quantum mechanics and condensed matter theory is the appearance of the nontrivial phase factor due to the adiabatic evolution of the eigenstates in the parameter space of  the Hamiltonian. The loop traced by the eigenstate is accompanied by a nontrivial, geometrical phase which is called as the Berry phase. The Berry phase is the holonomy associated to the $U(1)$ bundle. Topological and geometrical concepts also play an important role in quantum field theories. We would like to mention here, the axial or chiral anomaly,  which is the violation of the classically conserved chiral current at the quantum level. The expectation value of the divergence of the current results in the difference between the zero modes of opposite chirality which is the analytic index of the Weyl operator. Because of the Atiyah-Singer index theorem the analytic index is equal to the topological index which is given in terms of the Chern character (see \cite{be} and the references therein). 

Investigation of the physical properties with the help of topological invariants also become an efficient tool in condensed matter physics. One of the most remarkable achievements in this direction is the quantization of the Hall conductivity \cite{tknn} which is explained in terms of the first Chern number \cite{ass}. 

Within the homotopy theory the natural way for an investigation of the physical system is the use of fermionic Green's function $G(w,\bm{p})$ and its winding number. The winding number is given as an integer valued integral in terms of the Green's function and reflects some properties of the related Weyl particles \cite{v}. For example, in odd space-time dimensions effective field theory of the time reversal invariant topological insulators is the Chern-Simons action whose coefficient corresponds to the conductivity of the system. This coefficient, hence the conductivity is quantized as it is equal to the winding number of the fermion propagator \cite{qhz, dey}. In \cite{qhz, dey} it was also shown that the winding number is equal to the integral of the related Chern character given in terms of the Berry field strength. The same argument is valid in the case of the spin Hall effect \cite{del}. Recently, a possible condensed matter realization of Weyl Hamiltonian, namely Weyl semimetal is proposed \cite{wtvs} where the Berry phase reflects the nontrivial topological properties of its band structure. 

Chiral anomaly is the nonconservation of the classically conserved chiral current once the theory is quantized. However, recent research \cite{soy, sy} shows that a realization of chiral anomaly is possible in the context of semiclassical chiral kinetic theory by introducing the Berry gauge field induced through the diagonalization of the Weyl Hamiltonian. Moreover, a generalization for higher even dimensional spacetimes was achieved in \cite{ds}. In \cite{soy, sy}, the relation between the chiral magnetic effect and chiral anomaly in $3+1$ dimensions was also acquired. In \cite{de} both the chiral anomaly and the chiral magnetic effect is formulated in terms of differential forms in $5+1$ dimensions which can be easily generalized to higher dimensions. 

We would like to investigate the topological and geometrical concepts and the relations between them in the context of one particle Weyl systems including Berry gauge fields in any even $d+1$ dimensions. In momentum space, considering the $d+1$ dimensional Weyl Hamiltonian, we calculate the winding number of the fermion propagator which is a topological invariant \cite{v}.  The calculation results in the flux of a Dirac monopole field with a unit topological charge.  On the $d-1$ dimensional boundary of the momentum space, the winding number hence the charge can be expressed as the integral of the Chern character of the Berry gauge field which is the topological index. In Section 2 we give a general presentation of Berry gauge field and its field strength which are extracted from the diagonalization of the $d+1$ dimensional Weyl Hamiltonian. We define winding number of the 
Weyl fermion propagator in  $d+1$ dimensions. We first exhibit the formalism in $3+1$ and $5+1$ dimensions by calculating the related winding numbers and argue their connections with the Chern numbers and Dirac monopoles in Section 3 and Section 4, respectively. In Section 5, we carry our discussion to the higher dimensions. In the last section we summarize our conclusions and discuss the results.  

\section{Weyl Hamiltonian, the Berry Gauge Field and the Winding Number}

\subsection{Weyl Hamiltonian and the Berry Gauge Field}
In even $d+1$ dimensional spacetime, the one particle Weyl Hamiltonian
\be
\label{hwg}
{\cal{H}}_\ssW=\bm{\Sigma}\cdot\bm{p},
\ee
is expressed in terms of the $d$ dimensional momentum vector $\bm{p}$ and the $2^{\frac{d-1}{2}}\times 2^{\frac{d-1}{2}}$ dimensional $\bm{\Sigma}$ matrices satisfying the relations $\{\Sigma_{\ssM},\Sigma_{\ssN}\}=2\delta_{\ssM\ssN}$ where $M, N=1,..,d$. The Weyl Hamiltonian (\ref{hwg}) is derived from the massless Dirac Hamiltonian,
$$
{\cal{H}}_{\scriptstyle{D}}=\bm{\alpha}\cdot\bm{p}, \quad \{\alpha_{\ssM},\alpha_{\ssN}\}=2\delta_{\ssM\ssN},
$$
which is block diagonal in the chiral representation of the $2^{\frac{d+1}{2}}\times 2^{\frac{d+1}{2}}$  dimensional $\bm{\alpha}$ matrices. The Weyl equation is solved by the eigenvectors $|\psi_{\pm}^{(\alpha)}(\bm{p})\rangle$ as,
$$
{\cal{H}}_\ssW|\psi_{\pm}^{(\alpha)}(\bm{p})\rangle=\pm p|\psi_{\pm}^{(\alpha)}(\bm{p})\rangle,
$$
where $\alpha, \beta...=1,..,\frac{d-1}{2}$ indicating the $\frac{d-1}{2}$ fold degeneracy of the each eigenvalue $\pm p= \pm |\bm{p}|$. One can find a unitary matrix $U$ which diagonalizes (\ref{hwg}) as
\be
\label{diag}
U{\cal{H}}_{\ssW} U^\dagger={\rm{diag}}(p,-p)=p({\cal{I}}^+-{\cal{I}}^-).
\ee
${\cal{I}}^+$ and ${\cal{I}}^-$ are $2^{\frac{d-1}{2}}\times 2^{\frac{d-1}{2}}$ dimensional matrices projecting onto the positive and negative energy subspaces respectively:
\be
\label{im}
{\cal{I}}^+=\begin{pmatrix} 1 & 0 \\ 0 & 0 \end{pmatrix}, \quad  {\cal{I}}^-=\begin{pmatrix} 0 & 0 \\ 0 & 1 \end{pmatrix}.
\ee
By means of $U$ the Berry gauge field is introduced as 
\be
\label{dbgu}
{\cal{\bm{A}}}=i{\cal{I}}^+U\partial_{\bm{p}}U^\dagger {\cal{I}}^+.
\ee  
(\ref{dbgu}) is Abelian for $3+1$ dimensional spacetime however as a result of the $\frac{d-1}{2}$ fold degeneracy, it will be  non-Abelian in higher dimensions. Equivalently, the Berry gauge field also can be written in terms of the positive energy solutions:
$$
{\cal{A}}^{\alpha\beta}_\ssM=i\langle\psi^{\alpha}_+|\partial_{p_{\ssM}}|\psi^{\beta}_+\rangle.
$$
The related field strength is defined as
\be
\label{dbfs}
{\cal{G}}^{\alpha\beta}_{\ssM\ssN}=\partial_\ssM{\cal{A}}^{\alpha\beta}_\ssN-\partial_\ssN{\cal{A}}^{\alpha\beta}_\ssM-i[{\cal{A}}_\ssM,{\cal{A}}_\ssN]^{\alpha\beta},
\ee
where we used the shorthand notation $\partial_\ssM\equiv\partial_{p_\ssM}$.

\subsection {Winding Number and the Weyl Hamiltonian}

The $d+1$ dimensional winding number $C_d$ is an integer valued topological invariant which is defined in the momentum space of the system as an integral in terms of the Green's function $G(w,\bm{p})$ and its inverse $G^{-1}(w,\bm{p})$ \cite{v}:
\be
\label{wng}
C_d=N_d\int{d^dp\ dw\ \epsilon^{\mu\nu..\rho}\Tr[(G\partial_{p_\mu}G^{-1})(G\partial_{p_\nu}G^{-1})...(G\partial_{p_\rho}G^{-1})]}.
\ee 
Here, $\mu ,\nu=0,...,d$ and $\epsilon^{\mu\nu..\rho}$ is the $d+1$ dimensional totally antisymmetric Levi-Civita tensor. $N_d$ is the normalization constant which depends on the dimension $d$. (\ref{wng}) is not effected under the infinitesimal change $G\to G+\delta G$ which reflects its topological invariance. This property becomes invaluable if there exists a physical quantity corresponding to (\ref{wng}).

In order to construct $G(w,\bm{p})$ we invert the relation (\ref{diag})
\be
\label{dpog}
{\cal{H}}_\ssW=p(P^+-P^-),\quad P^{\pm}=U^{\dagger}{\cal{I}}^{\pm}U,
\ee
where $P^{\pm}$ are the projection operators:
\be
\label{ppog}
P^++P^-=1,\quad P^{\pm}P^{\mp}=0,\quad P^{\pm}P^{\pm}=P^{\pm}.
\ee
Now, it is possible to express $G(w,\bm{p})$ and its inverse $G^{-1}(w,\bm{p})$ by means of $P^\pm$ as
\be
\label{dgfg}
G(w,\bm{p})=\frac{P^+}{w-p}+\frac{P^-}{w+p}, \quad G^{-1}(w,\bm{p})=w-p(P^+-P^-).
\ee
The derivatives of $G^{-1}(w,\bm{p})$ with respect to $(w, \bm{p})$ can be calculated as
\be
\label{deriv}
\frac{\partial G^{-1}}{\partial w }=1,\quad \frac{\partial G^{-1}}{\partial{p_\ssM }}=-\left( \frac{p^\ssM}{p}(P^+-P^-)+p\partial_{\ssM}(P^+-P^-)\right).
\ee

In the next sections we will discuss the $3+1$ and $5+1$ dimensional cases explicitly.

\section{$3+1$ Dimensional Weyl Hamiltonian and the Winding Number}

In $3+1$ dimensional spacetime the Weyl Hamiltonian ${\cal{H}}^3_{\ssW}$ is given in terms of the Pauli spin matrices $\sigma_i$:
\be
\label{hw3}
{\cal{H}}^3_{\ssW}=\bm{\sigma}\cdot{\bm{p}}
=\begin{pmatrix} p_3 & p_1-ip_2\\ p_1+ip_2 & -p_3 \end{pmatrix}.
\ee
By means of the $2\times 2$ unitary matrix
\be
\label{u3}
U=\begin{pmatrix} N_+ & \frac{N_+(p-p_3)}{p_1+ip_2} \\ N_- & \frac{-N_-(p+p_3)}{p_1+ip_2} \end{pmatrix}, \quad N_+={\scriptstyle{\sqrt{\left( \frac{p+p_3}{2p}\right)}}},   N_-={\scriptstyle{\sqrt{\left( \frac{p-p_3}{2p}\right)}}}\ ,
\ee  
one can diagonalize the $3+1$ dimensional Weyl Hamiltonian (\ref{hw3}) as
$$
U{\cal{H}}^3_\ssW U^{\dagger}={\rm{diag}}(p,-p)=p({\cal{I}}^+-{\cal{I}}^-),
$$
in which ${\cal{I}}^+$ and ${\cal{I}}^-$ are defined as in (\ref{im}). The matrix (\ref{u3}) is constructed with the solutions $ |\psi_\pm(p)\rangle$  of the momentum space eigenvalue equation 
$$
{\cal{H}}^3_{\ssW} |\psi_\pm\rangle=\pm p|\psi_\pm\rangle,
$$
as 
$$
U=\big(|\psi_+(p)\rangle\ |\psi_-(p)\rangle\big)^\dagger.
$$
The spectral decomposition of (\ref{hw3}) is given in terms of the projection operators $P^\pm$ by inverting the relation (\ref{diag}):
$$
{\cal{H}}^3_{\ssW}=pU^\dagger({\cal{I}}^+-{\cal{I}}^-)U=p(P^+-P^-).
$$
$P^+$ can be calculated explicitly by (\ref{im}) and (\ref{u3}) as
$$
P^+=\frac{1}{2p}\begin{pmatrix} p+p_3 & p_1-ip_2 \\ p_1+ip_2 & p-p_3 \end{pmatrix}.
$$
Utilizing the definitions (\ref{dpog}) and (\ref{ppog}) it is possible to express $P^+$ by means of ${\cal{H}}^3_\ssW$:
\be
\label{p3}
P^+=\frac{1}{2}(\frac{{\cal{H}}^3_\ssW}{p}+1). 
\ee
Although we have presented the explicit forms of the unitary matrix $U$ (\ref{u3}) and the projection operator $P^+$ for completeness, in order to calculate the winding number we only need the definition of $P^+$ in terms of the Weyl Hamiltonian as in (\ref{p3}). Recalling (\ref{wng}), the $3+1$ dimensional winding number ${\cal{C}}_3$ is calculated by using (\ref{dgfg}) and (\ref{deriv}) as
\begin{eqnarray}
\label{dwn3}
{\cal{C}}_3&=&\frac{1}{8\pi^2}\int{ d^3p\ dw\ \epsilon^{\mu\nu\rho\sigma}\Tr[(G\partial_{\mu}G^{-1})(G\partial_{\nu}G^{-1})(G\partial_{\rho}G^{-1})(G\partial_{\sigma}G^{-1})]}\nonumber\\
 &=& \frac{1}{2\pi^2}\int{d^3p\ dw\ \epsilon^{abc}\Tr[(G^2\partial_aG^{-1})(G\partial_bG^{-1})(G\partial_cG^{-1})]} ,
\end{eqnarray}
where $\mu, \nu ...=0,..,3$, $a,b,c=1,2,3$ and $\Tr$ denotes the trace over the spin indices. It is straightforward to observe that the quadratic and the cubic terms in $p_a$ vanish due to the antisymmetry of $\epsilon^{\mu\nu\rho\sigma}$. A careful investigation shows that the terms linear in $p_a$ also give a vanishing contribution after the $w$ integration. Hence performing the $w$ integral and using the properties (\ref{ppog}), the winding number (\ref{dwn3}) is found to be:
\be
\label{cs3}
{\cal{C}}_3=-\frac{i}{2\pi}\int{d^3p\ \epsilon^{abc}\Tr[\partial_aP^+\partial_bP^+\partial_cP^+]}.
\ee
As $P^+$ is a $2\times 2$ matrix, (\ref{cs3}) does not vanish under the antisymmetry of the Levi -Civita tensor. However, the integrand is a total derivative,
$$
\int{d^3p\ \epsilon^{abc}\Tr[\partial_aP^+\partial_bP^+\partial_cP^+]}=\int{d^3p\ \bm{\nabla}\cdot \bm{K}_3}, 
$$
where we introduced,
\be
\label{k3}
K_3^a=\epsilon^{abc} \Tr[P^+\partial_bP^+\partial_cP^+]. 
\ee
Making use of (\ref{p3}) in (\ref{k3}) provides $K_3^a$ in the simple form, 
\begin{eqnarray}
\label{cp3}
K_3^a&=&\frac{1}{(2p)^3}\epsilon^{abc}\Tr[{\cal{H}}^3_\ssW(\partial_b{\cal{H}}^3_\ssW)(\partial_c{\cal{H}}^3_\ssW)] \nonumber \\
&=&\frac{1}{(2p)^3}\epsilon^{abc}\Tr[\bm{\sigma}\cdot\bm{p}\sigma_b\sigma_c]\\
&=&\frac{ip^a}{(2p)^3}2!\Tr[1_{2\times 2}]\nonumber .
\end{eqnarray}   
We employed the $SU(2)$ algebra and trace properties of the Pauli spin matrices. Therefore we obtain $K^a_3$ as
$$
K^a_3=\frac{ip^a}{2p^3},
$$
which is the field of a monopole with a unit charge located at $p=0$, that is $\bm{b}_3=\frac{\bm{p}}{2p^3}$. Hence, we conclude that the winding number (\ref{dwn3}) is the divergence of the field $\bm{b}_3$:
$$
{\cal{C}}_3=\frac{1}{2\pi}\int{d^3p}\ \bm{\nabla}\cdot\bm{b}_3.
$$
Using $\bm{\nabla}\cdot\bm{b}_3=2\pi\delta^3(p)$, we observe that the winding number ${\cal{C}}_3$ is equal to the unit charge of the monopole: 
\be
\label{wnmf3}
{\cal{C}}_3=\frac{1}{2\pi}\int{d^3p\ \bm{\nabla}\cdot(\frac{\bm{p}}{2p^3})}=\frac{1}{2\pi}\int{d^3p\ \bm{\nabla}\cdot\bm{b}}_3=1.
\ee
(\ref{dwn3}) is a topological invariant, therefore this monopole possesses a topological origin. In \cite{v}, it was stated that the winding number (\ref{dwn3}) denotes the chirality of the particle. Thus we conclude that the chirality defines the strength of the monopole.

Yet (\ref{cp3}) deserves a closer look. Definition of the projection operators (\ref{dpog}) enables us to express $K_3^a$ by means of the diagonalization matrix $U$ as
$$
K_3^a=\epsilon^{abc}\Tr[{\cal{I}}^+(\partial_bU)(\partial_cU^\dagger){\cal{I}}^+].
$$
The Abelian Berry gauge field is computed either in terms of the positive energy solution $|\psi_+\rangle$ or in terms of $U$ (\ref{u3}) as
\be
\label{bgf3}
{\cal{A}}^a=i\langle\psi^+|\frac{\partial}{\partial p_a}|\psi^+\rangle=i{\cal{I}}^+U\frac{\partial}{\partial p_a}U^\dagger{\cal{I}}^+=\frac{\epsilon^{ab3}p_b}{2p(p+p_3)}.
\ee
The related Berry field strength is calculated as, 
$$
{\cal{G}}_{ab}=\partial_a{\cal{A}}_b-\partial_b{\cal{A}}_a=i{\cal{I}}^+\big((\partial_aU)(\partial_bU^\dagger)-(\partial_bU)(\partial_aU^\dagger)\big){\cal{I}}^+.
$$  
We conclude that
$$
K_3^a=\frac{1}{2i}\epsilon^{abc}{\cal{G}}_{bc},
$$
which reveals the relation between the winding number (\ref{dwn3}) and the Berry field strength:
$$
{\cal{C}}_3=-\frac{1}{4\pi}\int{d^3p\ \epsilon^{abc}\partial_a{\cal{G}}_{bc}}.
$$
It was known that Berry curvature yields a monopole field $\epsilon^{abc}{\cal{G}}_{bc}=-\frac{p^a}{p^3}$ that is located at $p=0$ and this field is responsible for the chiral magnetic effect denoted in \cite{soy, sy, de} and the semiclassical chiral anomaly in 3 + 1 dimensions \cite{soy, sy, ds, de}. Hence, we conclude that both phenomena are related to the topological invariant (\ref{dwn3}) which is calculated in the momentum space. Using differential forms, on the 2-sphere $S^2$ which is the boundary of the 3-ball $B^3$, this relation turns out to be,
\be
\label{bcn3}
{\cal{C}}_3=-\frac{1}{4\pi}\int_{B^3}{d^3p\ \epsilon^{abc}\partial_a{\cal{G}}_{bc}}=-\frac{1}{4\pi}\int_{S^2}{d^2p\ \epsilon^{\rm{b}\rm{c}}{\cal{G}}_{\rm{b}\rm{c}}},
\ee
where $\rm{b},\rm{c}$ represent the polar and the azimuthal angles $\theta , \phi$ respectively and ${\cal{G}}_{\theta\phi}=\frac{\sin\theta}{2}$. (\ref{bcn3}) is the topological first Chern number multiplied with minus one. Over a compact manifold like $S^2$ the integral of the Berry curvature (Chern character) has to be a quantized number and we calculate it as $-1$. This equivalence of the Chern number with the winding number was also stated in \cite{qhz,dey} in the context of massive Dirac Hamiltonian in odd spacetime dimensions. 

We would like to emphasize the gauge field structure of the Dirac monopole (\ref{wnmf3}). We define the 1-form gauge field ${\cal{B}}_3$ as
\be
\label{agf3}
{\cal{B}}_3={\cal{A}}_a dp^a.
\ee   
${\cal{A}}_a$ is the Berry gauge field given in (\ref{bgf3}). ${\cal{B}}_3$ is defined on the upper hemisphere of $S^2$ as it is singular on the negative $z$ axis. (\ref{bgf3}) is also equal to the gauge field of the $U(1)$ Dirac magnetic monopole in $\bm{R}^3$ where quantization of the magnetic charge is given in terms of the winding number of the gauge group.  However, in our case we are dealing with a monopole of a unit charge emerging from the diagonalization of the Weyl Hamiltonian (\ref{hw3}). We have found that the winding number of the fermion propagator and the Chern number are equal to this unit charge. The gauge group of ${\cal{B}}_3$ is $U(1)$ and the Chern number (\ref{bcn3}) is known as the winding number of the principal bundle $P(S^2, U(1))$.

\section{$5+1$ Dimensional Weyl Hamiltonian and the Winding Number}

The $5+1$ dimensional Weyl Hamiltonian ${\cal{H}}_\ssW^5$ is a $4\times 4$ matrix expressed in terms of the $5$ dimensional momentum vector $\bm{p}$ and the $3+1$ dimensional Weyl Hamiltonian ${\cal{H}}_\ssW^3$ given in (\ref{hw3}) as 
\be
\label{hw5}
{\cal{H}}_\ssW^5=\bm{\varSigma}\cdot\bm{p}
=\begin{pmatrix} {\cal{H}}_\ssW^3 & i(p_4+ip_5)\\
-i(p_4-ip_5) & -{\cal{H}}_\ssW^3
\end{pmatrix}.
\ee   
$\bm{\varSigma}$ matrices are extensions of the Pauli spin matrices to $5$ dimensions:
$$
\varSigma_{a}=
\begin{pmatrix}
\sigma_{a} & 0\\
0 & -\sigma_{a}
\end{pmatrix};\ {\scriptstyle a=1,2,3},\quad
\varSigma_4=
\begin{pmatrix}
0 & i\\
-i & 0
\end{pmatrix}, \quad
\varSigma_5=
\begin{pmatrix}
0 & -1\\
-1 & 0
\end{pmatrix}.
$$
One can diagonalize the $5+1$ dimensional Weyl Hamiltonian (\ref{hw5}) as
$$
U{\cal{H}}^5_\ssW U^{\dagger}=\rm{diag}(p,-p)=p({\cal{I}}^+-{\cal{I}}^-),
$$
where ${\cal{I}}^\pm$ are now $4\times 4$ matrices (\ref{im}) and the
unitary matrix $U$ is 
$$
U=\begin{pmatrix}
\frac{-iN_+^1(p_1+ip_2)}{p_4+ip_5} & \frac{-iN_+^1(p-p_3)}{p_4+ip_5} & 0 & N_+^1\\
-iN_+^2(p_4-ip_5) & 0 & N_+^2(p-p_3) & -N_+^2(p_1-ip_2)\\
\frac{-iN_-^1(p_1+ip_2)}{p_4+ip_5} & \frac{iN_-^1(p+p_3)}{p_4+ip_5} & 0 & N_-^1\\
 iN_-^2(p_4-ip_5)& 0 & N_-^2(p+p_3) & N_-^2(p_1-ip_2)\\ 
\end{pmatrix}.
$$
It is  constructed in terms of the $2$-fold degenerate eigenstates $|\psi_\pm^{({\alpha})}(p)\rangle$, $\alpha=1,2$,  of the Weyl equation
$$
{\cal{H}}^5_{\ssW} |\psi^{(\alpha)}_\pm\rangle=\pm p|\psi^{(\alpha)}_\pm\rangle,
$$  
 where the normalization constants are: 
$$
N_+^{(1)}=\sqrt{\left(\frac{p_4^2+p_5^2}{2p(p-p_3)}\right)},\quad N_+^{(2)}=\frac{1}{\sqrt{(2p(p-p_3))}},\quad N_-^{(1)}=\sqrt{\left(\frac{p_4^2+p_5^2}{2p(p+p_3)}\right)},\quad N_-^{(2)}=\frac{1}{\sqrt{(2p(p+p_3))}}.
$$
By inverting the diagonalization process (\ref{diag}), it is possible to express (\ref{hw5}) in terms of the projection operators (\ref{ppog}) as
$$
{\cal{H}}^5_{\ssW}=pU^\dagger({\cal{I}}^+-{\cal{I}}^-)U=p(P^+-P^-).
$$
The $4\times 4$ matrix $P^+$ is explicitly calculated to be
$$
P^+=\frac{1}{2p}
\begin{pmatrix}
p+p_3 & p_1-ip_2 & i(p_4+ip_5) & 0 \\
p_1+ip_2 & p-p_3 & 0 & i(p_4+ip_5) \\
-i(p_4-ip_5) & 0 & p-p_3 & -(p_1-ip_2)\\
0 & -i(p_4-ip_5) & -(p_1+ip_2) & p+p_3
\end{pmatrix}.
$$
Using (\ref{dpog}) and (\ref{ppog}) we can rewrite $P^+$ in terms of the $5+1$ dimensional Weyl Hamiltonian (\ref{hw5}):
\be
\label{p5}
P^+=\frac{1}{2}(\frac{{\cal{H}}^5_\ssW}{p}+1).
\ee 
The winding number in $5+1$ dimensions is defined as
\begin{eqnarray}
\label{dwn5}
{\cal{C}}_5&=&-\frac{i}{48\pi^3}\int{ d^5p\ dw\ \epsilon^{\mu\nu\rho\sigma\lambda\gamma}\Tr[(G\partial_{\mu}G^{-1})(G\partial_{\nu}G^{-1})(G\partial_{\rho}G^{-1})(G\partial_{\sigma}G^{-1})(G\partial_{\lambda}G^{-1})(G\partial_{\gamma}G^{-1})]}\nonumber\\
 &=& -\frac{i}{8\pi^3}\int{d^5p\ dw\ \epsilon^{ijklm}\Tr[(G^2\partial_iG^{-1})(G\partial_jG^{-1})(G\partial_kG^{-1})(G\partial_lG^{-1})(G\partial_mG^{-1})]}, 
\end{eqnarray}
where $\mu,\nu...=0,...,5$ and $i,j...=1,5$. We perform the $w$ integration and find
$$
{\cal{C}}_5=\frac{1}{8\pi^2}\int{d^5p\ \epsilon^{ijklm}\Tr[\partial_iP^+\partial_jP^+\partial_kP^+\partial_lP^+\partial_mP^+]}.
$$    
This can be written as the total derivative,
$$
{\cal{C}}_5=\frac{1}{8\pi^2}\int{d^5p\ \bm{\nabla}\cdot\bm{K}_5}, \quad K_5^i=\epsilon^{ijklm}\Tr[P^+\partial_jP^+\partial_kP^+\partial_lP^+\partial_mP^+].
$$
We use the definition (\ref{p5}) and convert $K^i_5$ into the simpler form:
\begin{eqnarray*}
K^i_5&=&\frac{1}{(2p)^5}\epsilon^{ijklm}\Tr[{\cal{H}}^5_\ssW(\partial_j{\cal{H}}^5_\ssW)(\partial_k{\cal{H}}^5_\ssW)(\partial_l{\cal{H}}^5_\ssW)(\partial_m{\cal{H}}^5_\ssW)]\nonumber \\
&=&\frac{1}{(2p)^5}\epsilon^{ijklm}\Tr[\bm{\varSigma}\cdot\bm{p}\varSigma_j\varSigma_k\varSigma_l\varSigma_m] \\
&=&\frac{p^i}{(2p)^5}4!\Tr[1_{4\times 4}]=6\frac{p^i}{2p^5}\nonumber.
\end{eqnarray*}
We observe that the winding number (\ref{dwn5}), similar to the $3+1$ dimensional case, yields the Dirac monopole $\bm{b}_5=\frac{\bm{p}}{2p^5}, \bm{\nabla}\cdot\bm{b}_5=\frac{4\pi^2}{3}\delta^5(p)$. Therefore, we conclude that  the value of (\ref{dwn5}) is equal to the unit charge of the monopole:
$$
{\cal{C}}_5=\frac{3}{4\pi^2}\int d^5p\ \bm{\nabla}\cdot\bm{b}_5=1.
$$
In \cite{de},  the same monopole field was evoked in terms of the $2 \times 2$ matrix Berry gauge field,
\be
\label{bgf5}
{\cal{A}}_i^{\alpha\beta}=i\langle\psi^{(\alpha)}_+|\frac{\partial}{\partial p^i}|\psi^{(\beta)}_+\rangle=i({\cal{I}}^+U\frac{\partial}{\partial p^i}U^\dagger{\cal{I}}^+)^{\alpha\beta},
\ee
and its curvature
$$
{\cal{G}}^{\alpha\beta}_{ij}=\partial_i{\cal{A}}^{\alpha\beta}_j-\partial_j{\cal{A}}^{\alpha\beta}_i-i[{\cal{A}}_i,{\cal{A}}_j]^{\alpha\beta},
$$
as
\be
\label{mf5}
\frac{1}{24}\epsilon^{ijklm}\Tr[{\cal{G}}_{jk}{\cal{G}}_{lm}]=-\frac{pi}{2p^5}.
\ee 
Like in the $3+1$ dimensional case, this monopole field is the source of the chiral magnetic effect \cite{de} and the chiral anomaly \cite{ds,de} in the semiclassical chiral kinetic theory. We can accomplish the relation between the Berry gauge field (\ref{bgf5}) and the winding number (\ref{dwn5}) by substituting the explicit form of the projection operators (\ref{dpog}) into $K_5^i$:
\begin{eqnarray*}
K_5^i=\epsilon^{ijklm}\Tr &[&{\cal{I}}^+\partial_jU\partial_kU^{\dagger}{\cal{I}}^+\partial_lU\partial_mU^{\dagger}{\cal{I}}^+ \\ 
&+&2{\cal{I}}^+U\partial_jU^{\dagger}{\cal{I}}^+U\partial_kU^{\dagger}{\cal{I}}^+\partial_lU\partial_mU^{\dagger}{\cal{I}}^+ \\
&+&{\cal{I}}^+U\partial_jU^{\dagger}{\cal{I}}^+U\partial_kU^{\dagger}{\cal{I}}^+U\partial_lU^{\dagger}{\cal{I}}^+U\partial_mU^{\dagger}{\cal{I}}^+]\\
&=&-\frac{1}{4}\epsilon^{ijklm}\Tr[{\cal{G}}_{jk}{\cal{G}}_{lm}].
\end{eqnarray*}
Hence the winding number (\ref{dwn5}) and the monopole field (\ref{mf5}) which is responsible for the chiral anomaly in kinetic theory are related:
\be
\label{wnc5}
{\cal{C}}_5=-\frac{1}{32\pi^2}\int{d^5p\ \partial_i\epsilon^{ijklm}\Tr[{\cal{G}}_{jk}{\cal{G}}_{lm}]}. 
\ee

On the other hand letting the domain of the integral (\ref{wnc5}) to be the 5-ball $B^5$ whose boundary is the 4-sphere $S^4$, ${\cal{C}}_5$ can be written as
$$
{\cal{C}}_5=-\frac{1}{32\pi^2}\int_{B^5}{d^5p\ \epsilon^{ijklm}\partial_i(\Tr[{\cal{G}}_{jk}{\cal{G}}_{lm}]})=-\frac{1}{32\pi^2}\int_{S^4}d^4p\ \epsilon^{\rm{j}\rm{k}\rm{l}\rm{m}}\Tr[{\cal{G}}_{\rm{j}\rm{k}}{\cal{G}}_{\rm{l}\rm{m}}],
$$
where $\rm{j},\rm{k}...$ represent the angular coordinates on $S^4$. Therefore we find that the winding number is the negative of another topological invariant, namely second Chern number which is the integral of the second Chern character over $S^4$.  

As we deal with the Dirac monopole, we would like to explore its gauge field structure. In differential form language writing the 2-form Berry field strength as ${\cal{G}}=\frac{1}{2}{\cal{G}}_{ij}dp^i\w dp^j$ and recalling (\ref{mf5}) we define the Abelian 3-form antisymmetric gauge field ${\cal{B}}_5$ \cite{n} as  
\begin{eqnarray*}
\Tr[{\cal{G}}{\cal{G}}]=d{\cal{B}}_5.
\end{eqnarray*}
${\cal{B}}_5$ can be written explicitly as 
\be
\label{agf5}
{\cal{B}}_5=\Tr[{\cal{A}}d{\cal{A}}-\frac{2i}{3}{\cal{A}}^3],
\ee
where ${\cal{A}}$ is the Berry gauge field (\ref{bgf5}). Note that (\ref{agf5}) is in the form of Chern Simons Lagrangian. In its components Abelian rank-3 antisymmetric gauge field (\ref{agf5}) can be written as,
$$
{\cal{B}}^{ijk}_5=\Tr[{\cal{A}}^i\partial^j{\cal{A}}^k-\frac{2i}{3}{\cal{A}}^i{\cal{A}}^j{\cal{A}}^k].
$$

In the next section beginning with the  $d+1$ dimensional winding number of the fermion propagator ${\cal{C}}_d$, we will generalize the arguments that are presented in  Section 3 and Section 4. 

\section{$d+1$ Dimensional Weyl Hamiltonian and the Winding Number}
In the $d+1$ dimensional spacetime where $d+1$ is even, each eigenvalue $(p,-p)$ of the Weyl Hamiltonian (\ref{hwg}) is $\frac{d-1}{2}$ fold degenerate and in principle the corresponding eigenstates 
$$
|\psi_+^{(1)}\rangle,...,|\psi_+^{({\scriptscriptstyle{{\frac{d-1}{2}}}})}\rangle,\quad |\psi^{(1)}_-\rangle,...,|\psi_-^{({\scriptstyle{\frac{d-1}{2}}})}\rangle,
$$
can be computed. One can diagonalize (\ref{hwg}) by means of the unitary matrix $U$ which is constructed in terms of the solutions $|\psi_\pm^{\alpha}\rangle$ as
$$
U=\big(|\psi_+^{(1)}\rangle... |\psi_-^{({\scriptstyle{\frac{d-1}{2}}})}\rangle\big)^\dagger .
$$
With an appropriate choice of the normalization constant $N_d$, we write (\ref{wng}) as \footnote{Actually, ${\cal{C}}_d$ should be multiplied with $-1$ for the $3+1$ dimensional case in order to cancel the minus factor which will appear in the $K_3^M$ (\ref{cpd}). Because of this minus sign, in $3+1$ dimensions, we calculated the Chern number as $-1$ which is different from the general construction (\ref{nn}). This sign confusion is artificial in the sense that it is due to our choice of ${\cal{H}}^3_{\ssW}=\bm{\sigma}\cdot\bm{p}$ which is the conventional Weyl Hamiltonian used in the literature. If the Dirac matrices were constructed starting from the $1+1$ dimensions in chiral basis, there would not be this sign ambiguity. We must emphasize that the general definition of the winding number do not distinguish between the dimensions.}
$$
{\cal{C}}_{d}=\frac{i^{\frac{d+1}{2}}2^{\frac{3d-5}{2}}d}{\pi (d+1)! {d-1 \choose \frac{d+1}{2}}{\rm{Vol}}(S^{d-1}) }\int{d^dp\ dw\ \epsilon^{\mu\nu..\rho}\Tr[(G\partial_{p_\mu}G^{-1})(G\partial_{p_\nu}G^{-1})...(G\partial_{p_\rho}G^{-1})]}.
$$
Using the properties of the projection operators (\ref{ppog}) and the definitions (\ref{dgfg}), (\ref{deriv}) one can perform the $w$ integration and obtain,
\begin{eqnarray}
\label{cdk}
{\cal{C}}_{d}&=&\frac{i^{\frac{d+1}{2}}2^{\frac{3d-5}{2}}}{\pi (d-1)! {d-1 \choose \frac{d+1}{2}}{\rm{Vol}}(S^{d-1}) }\int{d^dp\ dw\ \epsilon^{\ssM \ssN ... \ssR}\Tr[G^2(\partial_\ssM G^{-1})(G\partial_\ssN G^{-1})...(G\partial_\ssR G^{-1})]}\nonumber\\
&=&\frac{-2(-2i)^{\frac{d-1}{2}}}{(d-1)! {\rm{Vol}}(S^{d-1})}\int{d^dp\ \epsilon^{\ssM ...\ssR}\Tr[(\partial_\ssM P^+)...(\partial_\ssR P^+)]}\\
&=&\frac{-2(-2i)^{\frac{d-1}{2}}}{(d-1)! {\rm{Vol}}(S^{d-1})}\int d^dp\ {\bm{\nabla}\cdot\bm{K}_d}\nonumber
\end{eqnarray}
where the italic capital letters denote $M ,N ,..,R=1,..,d$  and $\epsilon^{\ssM ...\ssR}$ is the $d$ dimensional totally antisymmetric tensor.  We defined $\bm{K}_d$ in terms of the projection operators as
$$
K_d^\ssM=\epsilon^{\ssM \ssN...\ssR}\Tr[ P^+(\partial_\ssN P^+)...(\partial_\ssR P^+)].
$$
It leads to the Dirac monopole in the $d$ dimensional momentum space:
\begin{eqnarray}
\label{cpd}
K_d^\ssM&=&\frac{1}{(2p)^d}\epsilon^{\ssM \ssN...\ssR}\Tr[{\cal{H}}_\ssW(\partial_\ssN{\cal{H}}_\ssW)...(\partial_\ssR{\cal{H}}_\ssW)]\nonumber \\
&=&\frac{1}{(2p)^d}\epsilon^{\ssM \ssN...\ssR}\Tr[\bm{\Sigma}\cdot\bm{p}\Sigma_\ssN...\Sigma_\ssR]\\
&=&-(d-1)!(\frac{i}{2})^{\frac{d-1}{2}}\frac{p^\ssM}{2p^d}.\nonumber
\end{eqnarray} 
Therefore for all spacetime dimensions considered, the winding number (\ref{wng}) is associated to the Dirac monopole,
\be
\label{mfd}
\bm{b}_d=\frac{\bm{p}}{2p^d},\  \bm{\nabla}\cdot\bm{b}_d=\frac{{\rm{Vol}}(S^{d-1})}{2}\delta^d(p).
\ee
Using the definition of the $d$ dimensional monopole field (\ref{mfd}), we can calculate (\ref{cdk}) as
\be
\label{wnmfd}
{\cal{C}}_{d}= \frac{2}{{\rm{Vol}}(S^{d-1})}\int_{B^d}{d^dp\ \bm{\nabla}\cdot\bm{b}_d}=1.
\ee
Thus the winding number ${\cal{C}}_d$ is equal to the unit charge of the $d$ dimensional Dirac monopole.

In order to display the general relation between the winding number and the Chern number, we point out that the winding number can be written as an integral over the $d$ dimensional ball $B^d$ as
\begin{eqnarray*}
{\cal{C}}_d&=&\frac{-2(-1)^{\frac{d-1}{2}}}{{\rm{Vol}}(S^{d-1})(d-1)!}\int_{B^d}d^dp\ \epsilon^{\ssM \ssN\ssR...\ssS\ssT}\partial_\ssM \Tr [\overbrace{{\cal{G}}_{\ssN\ssR}...{\cal{G}}_{\ssS\ssT}}^{\text{(d-1)/2 times}}]\\
&=&\frac{2(-1)^{\frac{d+1}{2}}}{{\rm{Vol}}(S^{d-1})(d-1)!}\int_{S^{d-1}}d^{d-1}p\ \epsilon^{\rm{\ssN\ssR...\ssS\ssT}}\Tr[\overbrace{{\cal{G}}_{\rm\ssN\ssR}...{\cal{G}}_{\rm{\ssS\ssT}}}^{\text{(d-1)/2 times}}],
\end{eqnarray*} 
where ${\cal{G}}_{\ssN\ssR}$ is the Berry field strength (\ref{dbfs}) and the letters $\rm{M,N...}$ represents the angular coordinates of the $d-1$ dimensional sphere $S^{d-1}$.
We observe that the monopole field (\ref{mfd}) which is obtained from the winding number can be expressed by means of the Berry curvature as
\be
\label{mfg}
b_d^\ssM=\frac{(-1)^{\frac{d+1}{2}}}{(d-1)!}\epsilon^{\ssM \ssN\ssR...\ssS\ssT} \Tr[\overbrace{{\cal{G}}_{\ssN\ssR}...{\cal{G}}_{\ssS\ssT}}^{\text{(d-1)/2 times}}]. 
\ee
(\ref{mfg}) is in accord with the previous results and it demonstrates that the monopole which is responsible for the chiral anomaly in chiral kinetic theory is the same with the one appearing in the winding number (\ref{mfd}). Since we can define the winding number (\ref{wng}) for even $d+1$ dimensions, we have proven the existence of semiclassical chiral anomaly for all even dimensions. Let $d=2n+1$ where $n$ is a positive integer. The volume of the $d-1$ dimensional sphere is given in terms of $\Gamma$ function as $\frac{2\pi^{d/2}}{\Gamma(d/2)}$. Using this, we find
\begin{eqnarray}
\label{bcnd}
{\cal{C}}_{2n+1}&=&\frac{\Gamma (n+\frac{1}{2})(-1)^{n+1}}{\pi^{n+\frac{1}{2}}\Gamma(2n+1)}\int_{S^{2n}}d^{2n}p\ \epsilon^{\rm{\ssN\ssR...\ssS\ssT}}\Tr[\overbrace{{\cal{G}}_{{\rm\ssN\ssR}}...{\cal{G}}_{{\rm\ssS\ssT}}}^{n\  times}]\nonumber \\
&=&\frac{(-1)^{n+1}}{(4\pi)^n n!}\int_{S^{2n}}d^{2n}p\ \epsilon^{\rm{\ssN\ssR...\ssS\ssT}}\Tr[\overbrace{{\cal{G}}_{\rm{\ssN\ssR}}...{\cal{G}}_{\rm{\ssS\ssT}}}^{n\  times}]=(-1)^{n+1}{\cal{N}}_n.
\end{eqnarray} 
${\cal{N}}_n$ is the $n^{th}$ Chern number defined as the integral of the $n^{th}$ Chern character. Comparing it with (\ref{wnmfd}) we conclude that, 
\be
\label{nn}
{\cal{N}}_n=(-1)^{n+1},
\ee
for all $n$. Up to a normalization, (\ref{nn}) is in accord with the spin Chern number computed in \cite{ds}. (\ref{bcnd}) shows that the winding number ${\cal{C}}_{2n+1}$ which is equal to the charge of the Dirac monopole, is actually the topological index given by the integral of the Berry curvature (\ref{dbfs}).

Lastly, we will investigate the gauge field structure of the $d$ dimensional Dirac monopole (\ref{mfd}).  A generalization of the Dirac magnetic monopoles to all dimensions by means of the antisymmetric tensor gauge fields was considered in \cite{n}. Utilizing the differential forms, we define the Abelian antisymmetric tensor gauge field ${\cal{B}}_{2n+1}$ as
$$
\Tr[{\cal{G}}^n]=d{\cal{B}}_{2n+1}=d\Tr[L_{\scriptscriptstyle{C}\ssS}^{2n-1}],
$$
where $L_{\scriptscriptstyle{C}\ssS}^{2n-1}$ is the $2n-1$ dimensional Chern-Simons Lagrangian. In its components we can write ${\cal{B}}_{2n+1}$ as
\be
\label{agfd}
{\cal{B}}_{2n+1}={\cal{B}}_{2n+1}^{\ssM...\ssR}\overbrace{dp_\ssM...dp_\ssR}^{2n-1\  times}=\epsilon^{\ssM\ssN\ssR...\ssS\ssT}\Tr[{\cal{A}}_\ssM\overbrace{{\cal{G}}_{\ssN\ssR}...{\cal{G}}_{\ssS\ssT}}^{n-1\ times}]\ d^{2n-1}p
\ee
which is in accord with (\ref{agf3}) and (\ref{agf5}) where ${\cal{A}}_\ssM$ is the Berry gauge field (\ref{dbgu}). We would like to emphasize that as (\ref{agfd}) is the field of the Dirac monopole, it is not defined globally on $S^{2n}$.

\section{Discussions}

Diagonalizing the $2n+2$ dimensional Weyl Hamiltonian (\ref{hwg}), we define the Berry gauge field whose gauge group is $U(n)$. The related holonomy
is given as the integral of the Chern character over the compact space $S^{2n}$. Both in $3+1$ dimensions and in $5+1$ dimensions we calculated explicitly the fermionic winding numbers and proved that they are equal to the unit monopole charge which emerges in momentum space. We showed that the winding number also can be stated as the integral of the Chern character, the topological invariant Chern number. The Chern character which is expressed via the Berry field strength (\ref{dbfs}) represents the nontrivial topological properties of the fiber bundle $(S^{2n},U(n))$. We observed that the winding numbers, monopole charges and Chern numbers are based on the same topological origin. We classified the gauge field structure of the Dirac monopole as $2n-1$ rank antisymmetric tensor gauge field as in \cite{n}.

In quantum field theory the chiral anomaly is the contribution of different chirality zero-modes to the measure of the path integral under the chiral transformation. The anomaly term is precisely equal to the analytical index of the Dirac operator which is projected to the positive chirality subspace. Due to the Atiyah-Singer index theorem, the analytical index is also topological in the sense that it can be given as the integral of the Chern class over the compact manifolds. On the other hand, in \cite{soy, sy, ds, de} it was argued that chiral anomaly can be achieved at the semiclassical level via the chiral kinetic theory by introducing the Berry gauge field. In the context of the one particle Weyl Hamiltonians, in all even dimensions we showed the existence of the Dirac monopoles on the degeneracy points, that is $p=0$. This monopole field acts as a source for the semiclassical anomaly. We observed that the topological invariant winding number (\ref{wng}) is equal to the charge of the monopole and according to \cite{v} this winding number results in the chirality of the Weyl particle. It is interesting but may be not surprising that even at the semiclassical level the chirality which shows up itself as a monopole located at the zero momentum degeneracy point is responsible for the chiral anomaly. Besides, this topological structure is also determined by the integral of the Chern character of the Berry curvature which is the topological index. We have done a topological and geometrical analysis of the semiclassical chiral anomaly and concluded that the concepts like the topological charges, fiber bundle theory, index theorems etc. also play an important role in the semiclassical theory. 

We would like to mention also that the $3+1$ dimensional Weyl Hamiltonian (\ref{hw3}) was argued to be the effective low energy Hamiltonian for the Weyl semimetals. In \cite{wtvs} it was noted that the stability of the Weyl semimetallic phase is related to the Chern number which is calculated in terms of the Berry gauge field (\ref{bcn3}). In this work, we show that this stability can be determined in terms of the another topological invariant, the winding number (\ref{dwn3}).

\subsection*{Acknowledgment}
I would like to thank \"{O}mer F. Dayi for fruitful discussions.

\newpage

\newcommand{\PRL}{Phys. Rev. Lett. }
\newcommand{\PRB}{Phys. Rev. B }
\newcommand{\PRD}{Phys. Rev. D }

\end{document}